# Hydrodynamics of countercurrent flows in a structured packed column: effects of initial wetting and dynamic contact angle

Rajesh Kumar Singh[*], Jie Bao, Chao Wang, Yucheng Fu and Zhijie Xu

Pacific Northwest National Laboratory, Richland, WA 99352, USA

*Corresponding Author: rajesh.singh@pnnl.gov; rajeshsingh.175@gmail.com

**ABSTRACT**

Computational countercurrent flow investigation in the structured packed column is a multiscale problem. Multiphase flow studies using volume of fluid (VOF) method in the representative elementary unit (REU) of the packed column can insight into the local hydrodynamics such as interfacial area, film thickness, etc. The interfacial area dictates the mass transfer in absorption process and thereby overall efficiency of column. Impacts of solvent's physical properties, liquid loads and static contact angle (SCA) on the interfacial area were examined earlier. In the present study, the dynamic contact angle (DCA) was used to explore the impact of contact angle hysteresis on the interfacial area. DCA has more pronounced impact on the interfacial area (10%) for aqueous solvent of 0.10M Sodium hydroxide (NaOH). The interfacial area shows undulation and does not achieve the pseudo-steady state. In contrary, the interfacial area gets a net pseudo-steady value for the aqueous solvent having 40% monoethanolamine (MEA) by `weight. The wetting hysteresis was also explored via simulations conducted with initially dry and wetted sheets. For 0.10M NaOH aqueous solvent, the initially wetted sheets lead to slightly higher value of the interfacial area (10%) as compared to the initially dry sheets at the same liquid load and DCA. As expected, wetting hysteresis reduces with increasing liquid loads. On the other hand, wetting hysteresis is not significant for 40% MEA aqueous solvent which might be lower surface tension and higher viscosity. Overall, the effect of the dynamic contact angle is not pronounced as compared to those found in a flat surface.

1. **Introduction**

Carbon dioxide ($CO_2$) capture in the packed column has emerged as a prospective technology to mitigate the greenhouse gas emission from thermal power plants for alleviating the impacts of global warming [1-3]. The post-combustion carbon capture via absorption could be economical and efficient because it can be readily retrofitted to the existing power plant. The structured packings is preferred since it provides higher effective mass transfer area with minimum pressure drop across the column [4]. However, the technology for the carbon capture is at the elementary stage, and it needs careful performance evaluation as well as scalability at the large scale before industrial deployment [5]. The experimental investigations for exploring the impacts of various factors on the column efficiency could be quite expensive and tedious. In contrary, computational fluid dynamics (CFD) studies can provide economic and efficient insights into the flow to complement the experiments. Subsequently, CFD has attracted more attentions to explore the flow characteristics in the structured packing over the decades. Moreover, flow simulations in the structured packed column is a multiscale problem due to presence of a wide range of length scales [6]. The microscale flow studies explore the local hydrodynamics involving flow regimes, film thickness, interfacial area, etc. Further, these flow characteristics are dependent upon various parameters, such as, solvent properties, liquid and gas loads and packing surface characteristics (contact angle, roughness and texture, and initial wettings). In this study, VOF method is used to investigate the effects of dynamic contact angle and the initial surface condition of sheets on the interfacial area. An enhanced understanding of the interfacial area evolution with the variation of DCA would be useful in the operation and design of a packed column.

A number of three-dimensional (3-D) CFD investigations were conducted for single-phase [7], pseudo single-phase [8, 9] and two-phase flow [10-13] to explore the hydrodynamics in the structured packings. Multiphase flow simulations were conducted for liquid falling over an inclined plate to explain the wetting of sheet in the packed column [14, 15]. Note that the triangular channels of corrugated sheet comprise of inclined plates with smaller width (10–30 mm). Further, studies for the falling film over an inclined plate was also employed to derive the models for liquid holdup [16] and interfacial area [17] in structured packings. Later, the pseudo single-phase approach was employed to calculate wet pressure drop across the column [8, 9]. In these studies, film thickness was computed in 2D simulation to derive liquid holdup for further calculation of wet pressure drop.

As mentioned earlier, interfacial area is a key factor for absorption efficiency, and numerous experimental [18-20] and computational studies [10, 11, 18, 21] were performed to investigate the interfacial area in the packed columns. Most of the earlier experimental studies to calculate the effective area [19, 22, 23] were based on the indirect measurement via chemical absorption in the fast chemical reaction. Janzen et al. [24] used X-ray computer tomography (XCT) to explore the impact of viscosity on liquid hold-up and interfacial area in MellapakPlus 752.Y packing. Both increase with increased value of viscosity and solvent flow rates. In contrary, the experimental study of Nicolaiewsky et al [25] reported a decrease in wetted area with increased viscosity. Tsai et al. [26] showed that solvent viscosity has a negligible effect whereas surface tension marginally affects the interfacial area. Rizzuti and Brucato [27] showed non-monotonic variation of the interfacial area with viscosity.

Multiphase studies using the VOF method were focused on the wetting corrugated sheet, which was arranged similar to the packed column by Shojaee et al. [13] and Subramanian and Wozny [28]. Shojaee et al. [13] indirectly calculated interfacial area via the average value of film thickness and liquid holdup in Gempak 2A packing. Furthermore, flow simulations were conducted in the REU of the packed columns [10-12, 21, 29, 30]. REU is represented as a small section that can replicate the identical hydrodynamic behavior of packing unit (see Figure 1(a)). Consequently, REU approach in the CFD simulations has become popular since it requires minimum computational resource. Ataki and Bart [18] conducted flow simulations in REU of Rombopack packing to study the impact of solvent properties on wetting, and also derived a correlation for the interfacial area. Haroun et al. [11] also used the VOF method to investigate the hydrodynamics in REU of Mellapak 250.X. The effects of solvent flow rate and SCA on the interfacial area were investigated. Basden [21] also investigated the interfacial area and liquid holdup in the REU. The effects of various factors affecting the interfacial area were explored; however, attempted studies were scattered. Sebastia-Saez et al. [12] also investigated a meso-scale model based on REUs of the MontzPak

B1-250 packings to calculate the liquid holdup and the interfacial area. Recently, multiphase flow investigations [10] in the REU of Gempak 3A were extensively conducted to develop correlation for interfacial area and liquid holdup in a packed column. The correlations account for impacts of various factors such as solvent properties, liquid load, and static contact angle on the liquid holdup and interfacial area.

In addition to the solvent properties and flow rates, wetting in a packed column is also governed by the surface behavior of the packing material i.e., contact angle ($\gamma$) and surface texture [31]. The value of contact angle depends on solid–liquid systems in a specific environment [32]. For example, contact angle measured at the vapor-liquid equilibrium condition is smaller than that obtained under atmospheric conditions [33]. Furthermore, the surface texture of packing material tends to reduce the apparent contact angle as compare to that obtained for a smooth surface. For a homogenous series of liquids, the value of *cosγ* linearly varies with surface tension ($\sigma$) value at a given solid. Hence, a lower σ value can yield to a smaller value of contact angle [34]. Further, it is a critical factor that significantly affects the wetting ability of a solvent for a given solid. A lower value of contact angle leads to enhanced interfacial area [15, 35]. In addition, the surface texture promotes wetting in the packed column [36, 37] by reducing the value of apparent contact angle [38, 39]. Recent studies for falling rivulets over an inclined plate [15] and a single corrugated sheet [40] have shown that the impact of contact angle on interfacial area ties with the surface tension of the solvent. Effects of contact angle on the interfacial area were found [10, 35] to be pronounced for a solvent having lower surface tension value. However, most of the empirical effective area models [36, 41] assumed uniformly distributed wetted sheets and overlooked the influence of contact angle. Rocha et al. [42] and Gualito et al. [43] overestimate the impact of contact angle. They modified the Shi and Mersmann model [35] which was originally derived from the liquid falling over an inclined plate. Further, the pre-wetting of packing unit is generally performed before start-up of capture operation in the industry to promote the wetting, thereby enhancing the column efficiency. The structured packed column is prewetted via opening of the top valve and subsequently the circulation of solvent in the packed column for 15–30 min at the beginning of operation [19]. Previous studies were adhered to an initially dry sheet and overlooked the impact of initial wetting.

The wetting of the packing surface in the structure packed column is quite complex and is not yet fully understood. Furthermore, measurement of contact angle in a structured packed column is also difficult because of complex design and operating environment. Previous studies describing the effects of contact angle on the interfacial area were adhered to the static contact angle. The advancing contact angle reflects the wetting behavior of a liquid on a dry surface while the receding contact angle is a measure of the remaining solid-liquid interaction forces. The difference between advancing and receding contact angles ($\gamma_A - \gamma_R$) is referred as the contact angle hysteresis, and its value can be significant [44]. It is a consequence of surface imperfections or simply irreversible interaction of the contact liquid and solid [45]. DCA can provide an accurate description of the wetting and spreading process, and it can help to characterize surface wettability. To our knowledge, impact of the dynamic contact angle was not ever investigated earlier flow simulations in the structured packings. Indeed, previous studies [12, 46] also suggested to include dynamic contact angle on the sheets in the flow simulations for the prediction of local hydrodynamics in the structured packings. Accordingly, the present study investigates the impact of dynamic contact angle on the interfacial area in a packed column. In addition, the effect of the initial wetting of sheets on the interfacial area is also explored for closely industrial conditions.

In this paper, the countercurrent flows in the REU of Mellapak 250.Y packing is extensively studied using VOF method. Mathematical formulation is described in section 2. Next, problem setup, meshing and numerical scheme used in the flow simulation are briefly discussed. Results describing the impact of various parameters affecting the interfacial area are explained in section 4. The comparison of predicted results with existing experimental data for the effective area is presented. Effects of dynamic contact angle and initial wetting (dry vs wetted) are discussed. A summary is presented last.

2. **Mathematical formulations**

Multiphase flow simulations for the countercurrent flows in the REU of Mellapak 250.Y packing were performed using a commercial CFD code STAR-CCM+ [47]. The governing equations are as follows:

$$\nabla \cdot \mathbf{u} = 0, \quad (1)$$

$$\frac{\partial(\rho \mathbf{u})}{\partial t} + \nabla \cdot (\rho \mathbf{u}\mathbf{u}) = -\nabla p + \mu \nabla \cdot \left(\nabla \mathbf{u} + (\nabla \mathbf{u})^T\right) + \rho \mathbf{g} + \mathbf{F}, \quad (2)$$

Here, $\mathbf{u}$ is the velocity, $p$ is the pressure, $g$ is the gravity, and $\mathbf{F}$ is the interfacial surface tension force that produces a normal pressure jump across the interface. The terms $\rho$ and $\mu$ are phase average density and viscosity respectively, which is computed as:

$$\left. \begin{aligned} \rho &= \rho_g + \alpha(\rho_l - \rho_g) \\ \mu &= \mu_g + \alpha(\mu_l - \mu_g) \end{aligned} \right\} \quad (3)$$

here α is the volume fraction of the primary phase and suffices $l$ and $g$ denote the liquid and gas phase, respectively.

The conservation equations (1) and (2) are solved using the VOF method [48], which allows the entire flow field to be treated as a single phase. The interfacial force ($\mathbf{F}$) appearing at the gas–liquid interface is implemented by continuous surface force model (CSF) [49]:

$$\mathbf{F} = \sigma \frac{\rho \kappa \nabla \alpha}{\frac{1}{2}(\rho_g + \rho_l)} \quad (4)$$

here σ is the interfacial tension value, κ is the local curvature of the interface, and $\nabla \alpha$ is the gradient of the volume fraction representing the direction vector at the gas-liquid interface. The curvature κ is computed as the divergence of the unit normal ($\hat{n} = \nabla \alpha / |\nabla \alpha|$).

$$\kappa = \nabla \cdot \hat{n} \quad (5)$$

The interface where two fluids meet the wall is computed by wall adhesion and the contact angle (γ). The normal vector at the interface is adjusted near the wall by the following equation:

$$\hat{n} = \hat{n}_{wall} \cos \gamma + \hat{t}_{wall} \sin \gamma \quad (6)$$

here $\hat{n}_{wall}$ and $\hat{t}_{wall}$ are the unit vectors normal and tangential to the wall, respectively.

The interface between the phases is captured by solving addition transport equation (7) for α whose value varies from 0 to 1.

$$\frac{\partial \alpha}{\partial t} + \nabla \cdot \mathbf{u}\alpha + \nabla \cdot \mathbf{u}_c \alpha(1-\alpha) = 0 \qquad (7)$$

Here, $u_c$ is the sharpening velocity computed as $C \times |u| \frac{\nabla \alpha}{|\nabla \alpha|}$. $C$ (=0.25) is sharpening factor for reducing numerical diffusion. The last term in the equation (7) acts against the interface smearing and assures the smooth shape interface. Note that it only acts in the vicinity of interface and therefore does not significantly affect the solution in the rest of the flow domain.

## 2.1 Dynamic contact angle

Modeling of the contact angle is a challenging issue in the numerical treatment of solid–liquid interaction involved in the present studies. Young's equation [50] presents a relationship between the contact angle and the surface tension; however, in practice the observed contact angle is not equal to that defined by Young's equation [51]. The contact angle on the solid surface varies as the liquid front evolves because of the contact line motion during wetting and spreading (See Figure 2). It changes around its equilibrium value during the motion of a solvent [52] and commonly called dynamic contact angle. The dynamic contact angle, measured at the advancing ($\gamma_A$) or receding ($\gamma_R$) edge of the rivulet, differs from the static contact angle [53]. The shear stress singularity arises at the triple phase contact line from the no-slip condition at the wall. Finding a relationship between the dynamic contact angle and contact line velocity ($u_{CL}$) is a key aspect in dealing with wetting phenomenon. Modeling the wetting phenomenon accounting for the change in advancing and receding contact angles leads to a singular force term in the momentum equation. The Kistler model [54] for dynamic contact angle was chosen, which is suitable for both inertia and capillary dominated flows. The empirical correlation is based on the capillary number and usage of the Hoffman function and computed as:

$$\gamma_k = f_{Hoff}\left[Ca + f_{Hoff}^{-1}(\gamma)\right] \qquad (8)$$

Here, Ca is capillary number ($Ca = \mu u_{CL}/\sigma$) based on $u_{CL}$ defined as:

$$u_{CL} = -(\mathbf{u_r} \cdot \hat{n}_t) \qquad (9)$$

$\mathbf{u_r}$ is the relative velocity of the fluid and the corresponding wall at the triple contact line. $\hat{n}_t$ is the unit vector in the tangential direction pointing in the same direction as the volume fraction gradient of the primary phase ($\nabla \alpha$).

The value of $\gamma$ is substituted in the equation (8) according to the sign of $u_{CL}$:

$$\gamma = \begin{cases} \gamma_A & \text{for } u_{CL} > 0 \\ \gamma_R & u_{CL} < 0 \end{cases} \qquad (10)$$

$f_{Hoff}^{-1}$ is the inverse Hoffmann function which is defined as:

$$f_{Hoff}(x) = \cos^{-1}\left[1 - 2\tanh\left\{5.16\left(\frac{x}{1+1.31x^{0.99}}\right)^{0.706}\right\}\right] \qquad (11)$$

A range for the equilibrium Capillary number ($Ca_E$) value is implemented to enhance the stability of the simulation. Within the specified range ($-Ca_E < Ca < Ca_E$), the resulting dynamic contact angle is blended with the equilibrium contact angle ($\gamma_E$).

$$\gamma_d = f\gamma_E + (1-f)\gamma_k \qquad (12)$$

The factor $f$ is computed as:

$$f = 0.5 + 0.5\cos\pi(Ca/Ca_E) \qquad (13)$$

As evident, the model requires several input variables to characterize the dynamic contact angle: $\gamma_A$, $\gamma_R$, $Ca_E$ and $\gamma_E$. The selection of these is discussed later.

### 2.2 Species transport

The single-fluid formulation for mass transfer between two phases developed by Haroun et al. [55, 56] was used, which is capable of simulating species transport across arbitrary interfaces morphology using interface capturing methods.

$$\frac{\partial C_i}{\partial t} + \nabla \cdot (\mathbf{u}C_i - D_i \nabla C_i) = S_i \qquad (14)$$

Here, $C_i$ and $D_i$ is the concertation and diffusivity of the $i^{th}$ species. $S_i$ is the species source term that accounts the production of $i^{th}$ species due to chemical reactions. The computation for the source term was followed by the method explained by Wang et al. [57].

$$S_i = r\alpha C_S C_{CO_2} \qquad (15)$$

Here, $r$ is reaction rate constant and $C_S$ is the concentration of solvent ($C_S \approx [OH^-]$ or $[RNH_2]$ for aqueous caustic solvent and MEA respectively). Note that the pre vious reactive multiphase flow studies by Wang et al. [57] were performed in a 2D domain mincing a wetted wall column. Because of the small 2-D domain, mesh resolution was not challenging. One needs extremely fine mesh near the interface in the reactive mass transfer phenomenon, which is not affordable in the 3-D simulation of structured packing. To overcome, the equation (15) was modified via introducing additional polynomial terms.

$$S_i = r\alpha^a (1-\alpha)^b C_S C_{CO_2} \qquad (16)$$

The additional factor was expected to overcome the impact of slightly coarser mesh near the interface. The value of the exponential factors $a$ and $b$ was calibrated against experimental data and found to $a = b = 1.6$.

## 2.3 Effective area via absorption of CO₂ in aqueous solvent

The effective mass transfer area in the packed column is computed via Danckwerts method [58], which is based on absorption of $CO_2$ into aqueous NaOH solvent. The rate of reaction is considered as instantaneous reaction [59]. Such condition, the physical mass transfer is negligible as compared to the corresponding value for reactive mass transfer [60]. The normalized effective mass transfer area ($A_{eff}$) in the packed columns is calculated as:

$$A_{eff} = \frac{u_g}{k'_g ZRT} \ln\left(\frac{C_{CO_2,in}}{C_{CO_2,out}}\right) \cdot \frac{1}{A_p} \qquad (17)$$

Here, $u_g$ is gas velocity, $Z$ is height of column, R is the molar gas constant, $T$ is the temperature and $A_p$ ($= 250 \, m^2/m^3$) is the specific area of packing. $k'_g$, the overall mass transfer coefficient, was computed using the correlation developed by Pohorecki and Moniuk [59]. The value of the $k_g'$ and $r$ used in the flow simulations are presented in Table 2 for both solvents and their detailed derivation can be found somewhere else [61].

The expression (17) for computing effective area was derived with assumptions such as same area of inlet and outlet, uniform gas velocity, etc. During the computation of effective area, flux of $CO_2$ was used instead of its concentration.

$$A_{eff} = \frac{u_g}{k_g' ZRT} \ln\left(\frac{J_{in}}{J_{out}}\right) \cdot \frac{1}{A_p} \tag{18}$$

Here, $J_{in} = C_{CO_2,in} \cdot u_{g,in} \cdot A_{in}$ and $J_{out} = C_{CO_2,out} \cdot u_{g,out} \cdot A_{out}$ are flux of the $CO_2$ at inlet and outlet of gas phase respectively.

3. **Problem setup and numerical scheme**

We conduct multiphase flow simulations for the computation of the effective area in the REU of Sulzer Mellapak 250.Y packing (Figure 1(a)). Similar to the previous studies [10, 21, 62], a model of the computational flow domain was created by perpendicular arrangement of two smooth corrugated sheets with 2 *mm* minimum gap between them. It consists of a single repeating unit in the lateral direction and three units in the flow direction (see Figure 3(a)). Design of the corrugated sheet is same as the design of Mellapak 250.Y (see Figure 1(b) and Table 1), and perforation was not included in the flow simulation. Previous studies by Green et al. [63] suggested that liquid films mostly cover holes at the majority of the operated liquid loads, and, eventually, it leads to minimal area loss. Therefore, results for sheets without perforation would not be significantly different from perforated ones. A schematic of the computational flow domain is presented in Figure 3(a). The solvent enters in the flow domain through its top of sheet and leaves from the bottom due to gravity (see Figure 3(a)). In the simulation, the sheet was set as no-slip walls with the static contact angle ($\gamma_S$). Both sides of the domain were specified as periodic boundary conditions that allow flow to cross and reenter in the flow domain through lateral boundaries (side walls). Symmetry boundary condition was used in the previous studies [12, 64] in the similar setup. The symmetric boundary condition constraints flow inside the REU, which can be seen in Figure 4 of reference [12]. Therefore, it might not be a realistic boundary condition for such flow simulations. The bottom of the flow domain was set to pressure outlet boundary with atmospheric condition. The uniform inlet velocity was specified at the liquid inlet and top of the flow domain. Positive (liquid inlet) and negative (at the top for gas) values of the velocity were specified to impose countercurrent flows. Similar boundary conditions for liquid and gas inlets were previously used in such flow simulations [11, 13].

**Table 1: Design of the corrugated sheet for Mellapak 250.Y packing**

| Parameters | Dimension |
|---|---|
| Corrugation angle ($\alpha$) | 45° |
| Corrugation base (B) | 26.70 mm |
| Corrugation height ($h$) | 12.00 mm |
| Corrugation side (s) | 17.00 mm |
| Specific Area ($a_p$) | 250 m²/m³ |
| Void Fraction ($\varepsilon$) | 0.96 |

Due to the complex model of the REU (in particular, corrugated sheets comprised of inclined triangular channels) modeling the flow domain is challenging. The model of flow domain was created in the 'Salome 8.3', opensource CAD software according to sheet's design presented in Table 1. The CAD model of the flow domain was further imported in the STAR-CCM+ for meshing and subsequent flow simulations. Meshing of the domain is one of the crucial steps affecting stability and accuracy of the results. To capture hydrodynamics and interphase mass transfer accurately and efficiently, flow domain was discretized with nonuniform mesh (Figure 3(b)). As shown in the exploded view of meshing of flow domain in Figure 3(b), the mesh near the corrugated sheets was created finer using prism layer meshing (similar to boundary layer mesh in other software) scheme. In this region, liquid film appears and interphase mass transfer occurs; consequently, it requires very fine mesh. Inflexion regions and the crest and valley of the triangular channels were also refined to capture the hydrodynamics. Therefore, a minimum grid size of 25 μm was specified in this region. Further, the gradient based trimmed mesh was used to discretize the reaming section of the flow domain. Subsequently, the size of mesh proximate to the prizm layer is smaller than the mesh in the center of domain (see in Figure 3(b)). Note the center of domain does not actively contribute in transport phenomenon of interest. Subsequently, 3.25 million cells in the flow domain were obtained after grid convergence tests. The flow simulations required very small-time step ($\Delta t \sim 10^{-6} - 10^{-5}\ sec$) to satisfy the Courant–Friedrichs–Levy (CFL) condition. Consequently, flow simulations were very expensive, e.g., a single simulation takes 4 –7 days wall time using 192 cores to achieve the pseudo steady state depending upon solvent properties and contact angle.

The unsteady flow simulations were conducted via implicit transient formulation that involves specified inner iterations in a time step to converge the solution. The coupling between the momentum and continuity equations is achieved with a predictor-corrector approach based on the Rhie and Chow interpolation method in conjunction with the SIMPLE (Semi-Implicit Method for Pressure-Linked Equations) algorithm [65] for pressure-velocity coupling. The second-order upwind scheme was used in the spatial discretization of all equations. The simulation employed an implicit solution scheme in conjunction with an algebraic multigrid method (AMG) for accelerating convergence of the solver. The high-resolution interface capturing (HRIC) was used in the discretization in interface, which is suitable for tracking sharp interfaces [66]. Convergence of the solution was assumed when the sum of normalized residual for each conservation equation was less than or equal to $10^{-5}$. In addition to that, simulations were terminated when the concentrations of the $CO_2$ at the outlet and inlet achieved pseudo steady state.

The multiphase flow studies were conducted with flue gas ($\rho_g$ =1.185 kg/m³ and $\mu_g$ =1.831×10⁻⁵ Pa.s), and aqueous solvents utilized in carbon capture. The solvents include aqueous solution of 0.10M NaOH (henceforth NaOH) and MEA at 0.40 concentrations by weight (40MEA). The effects of physical properties are presented in term of the Kapitza number (Ka) which is defined as $Ka = \sigma \left(\frac{\rho_l}{g\mu_l^4}\right)^{1/3}$. It is a dimensionless number [60] representative of the fluid properties only. A solvent having high surface tension and low viscosity results in higher Ka value. The flue gas with a given $CO_2$ concentration is fed from the bottom and gas is released from the top of the flow domain. Tsai et al. [36, 67] calculated the gas-side resistance using the correlation developed by Rocha et al. [17] for absorption of $SO_2$ into caustic solvents at four gas flow rates ($u_g = 0.6, 1.0, 1.5$ and $2.3$ m/sec ). The gas-side mass transfer resistance was found to be 1–2% of overall mass transfer resistance. Therefore, the gas-side mass transfer resistance can be neglected, and overall mass transfer would be approximately equal to the liquid-side mass transfer. Accordingly, gas velocity was chosen as 1.0 m/sec to ignore the gas-side mass transfer resistance. The physical properties of these solvents are presented in Table 2. The value of contact angle (γ) for each solvent was not available. It would be worthy to note that the $CO_2$ absorption is carried out in the prewetted column in which surfaces of the packing material are non-planar, textured, and heterogeneous. Because of the prewetted textured sheet, one can expect lower value of the contact angle. Subsequently, lower value of the contact was used in the validation of the flow simulation. Later, effects of static and dynamic contact angles are systemically studied.

**Table 2: Properties of solvents used in flow simulations**

| Solvent | $\mu_l$ (mPas) | $\rho_l$ (Kg/m³) | σ (mN/m) | Ka | r (m³/mol s) | $k'_g$ (mol/m² pa s) |
|---|---|---|---|---|---|---|
| NaOH (0.10M NaOH) | 0.97 | 997.0 | 72.80 | 3538 | 8.466 | 4.09×10⁻⁷ |
| 40MEA (40% MEA) | 3.791 | 1053 | 54.80 | 443 | 5.964 | 1.56×10⁻⁶ |
| | % = percentage by weight, M = molar concentration | | | | | |

## 4. Results and discussions

The results of flow simulations for the prediction of the interfacial area are discussed, which account for the effects of liquid and gas loads, contact angles (static and dynamic) and initial wetting of sheet. The predicted effective area is compared first with the experimental value of effective area for the Mellapak 250.Y. The CFD predicted shape of the dynamic rivulet (meandering) is also compared with the experimental observation. Further, effects of the liquid load, contact angle and initial wetting of sheets on the interfacial area is presented. The dynamic contact angle accounts the contact line velocity, advancing and receding contact angles.

### 4.1 Grid convergence test

The grid convergence test is a prerequisite for the numerical studies that aids to estimate the optimal number of cells in the flow domain for efficient and economic simulations. A coarser grid resolution can lead to unstable simulation as well as inaccurate result. While a very fine grids leads to large number of cells, and a very small-time step is required for stable and converged solution. Consequently, flow simulations become computationally very expensive. In this view, flow simulations were initially performed for 40MEA with DCA ($\gamma_A = 45°$, $\gamma_R = 30°$ and $\gamma_E = 40°$) at $q_L = 62 \, m^3/m^2h$ and six grid resolutions to obtain reasonable number of cells in the flow domain. The interface is defined as the iso-surface corresponds to $\alpha = 0.50$, which has already been adopted earlier for representing the gas/liquid interface [11, 18, 40]. The area of the interface ($A_I$) was computed for all cases, and further normalized by the specific area of the packing ($A_p = 250 \, m^2/m^3$) as $A_{In} = A_I/A_p$. The optimum number of cells was established based upon the value of $A_{In}$ at these grid resolutions. Figure 4 shows the temporal evaluation of $A_{In}$ at these cases. As expected, $A_{In}$ first increases with time, and then achieves a net pseudo-steady value for all cases. The pseudo-steady value of $A_{In}$ converges with increased number of cells, i.e. enhanced grid refinement. The intermediate grid resolution of 3.18 M cells was chosen for further simulations. Although there is a slight deviation in the value of $A_{In} (= 0.38)$ at this grid resolution, but the shape of interface does not significantly differ from that at 4.16 M cells ($A_{In} = 0.36$) (see inset of Figure 4). The discrepancy in the pseudo-steady value of $A_{In}$ is within the reasonable limit (~5%). The chosen grid resolution (3.18 M cells) took 3 days of wall clock time on 144 cores to achieve pseudo-steady state while the finest resolution (4.16 M cells) took almost a week.

**4.2    Comparison with experiments at different liquid loads**

Preliminary flow simulations were conducted to validate the numerical methods that reflects the extent of accuracy. Accordingly, predicted effective areas are first compared with those obtained from experiment for Mellapak 250.Y structured packings [67]. Tsai et al. [67] measured the effective area of Mellapak 250.Y packed column having 0.46 m and 0.427m outer and inner diameters, respectively. Aqueous NaOH was used as solvent and distributed via pressurized fractal distributor with 108 drip points/m². Flue gas was introduced from the bottom of packings in a range of velocity as 0.50–1.50 m/s. The column was prewetted via opening the top valve and subsequently circulation of solvent in the packed column for 15–30 min. The concentration of $CO_2$ at the inlet and outlet was monitored. The liquid load was incremented, and value of liquid load was maintained until $CO_2$ concentration at the outlet achieved steady state. Details of the experimental studies can be found in somewhere else [19, 67].

According to experiments, flow simulations were conducted for the gas velocity of 1 m/sec and aqueous 0.10M NaOH solvent at different liquid loads. The flue gas contains a concentration of $CO_2$ as 410 ppm. Due to pre-wetted textured sheet, one can expect lower value of the contact angle. Because of computing resource restriction, the flow simulations were restricted to a smooth surface. Note that surface texture/embossment has been reported to reduce the apparent contact angle, thereby promoting sheet wetting [38, 39]. It was implemented via the apparent/Wenzel contact angle in the flow simulation. Hence, lower value of the contact angle was used in the flow simulations for representing the experimental conditions. The effect of the dynamic contact angle on the wetting for a solid-liquid system having low equilibrium contact angle is found to be insignificant. Therefore, static contact angle was specified at the wall instead of dynamic contact angle. Moreover, the flow simulations were also validated for the dynamic contact angle and is presented in the next section. The flow simulations were run until a pseudo steady state was achieved regarding the constant value of outlet and inlet flux of $CO_2$ (see Figure 5 (a)). Further, the effective area was computed using the equation (18) for the gas velocity of 1.0 m/sec. Figure 5(b) shows the comparison between CFD predicted normalized effective mass transfer area ($A_{eff}$) and corresponding experimental value. As expected, $A_{eff}$ increases with increasing liquid loads for both methods. Further, $A_{eff}$ is observed to more than one at the higher liquid loads. It may be due to the presence of droplets and slander liquid threads in open space as well as the ripples at the gas-liquid interface for the liquid attached to the sheets. Combining contribution of these with smooth gas-liquid interface may give rise to larger value of the effective area ($A_{eff} > 1$). It was also shown by Bravo and Fair [68]. The CFD predicted effective area matches well with the corresponding one for experiments. A slight discrepancy is found at lower liquid load; otherwise, both show good matching.

The effective mass transfer area in the experiments was indirectly calculated via gas absorption. Note that the chemical reaction takes place at the gas-liquid interface. In VOF simulation, the gas-liquid interface is represented as the iso-surface corresponds to $\alpha = 0.50$ and a shape of the interface is shown in the inset of Figure 5(b). In this view, the interfacial area could be analogous to the effective mass transfer area.

## 4.3 Comparison with experiments for dynamic rivulets

Flow simulations were also validated for dynamic rivulet, particularly meandering rivulet falling over inclined plate. In our previous study [69] for the breakup of rivulet falling over an inclined plate, CFD prediction for the wetted area and the shape of the dynamic rivulets were found to compare well with the results of in-house experiments. In that work, experiments were carried out for water rivulet falling on a smooth plate which is 60° inclined to the horizontal. Water was introduced to the top of the plate by flowing over a 20 mm wide weir. Details of the experiment can be found elsewhere in [69]. A 1 cm square grid was placed underneath the plate for dimensional reference. Snapshots of the flow and a corresponding short video were recorded at different flow rates.

In line with experiments, flow simulations were conducted for water and air as working fluids. Water was falling on an inclined plate whereas air was considered as the stagnant phase. The Meandering rivulet emerges at low flow rate that significantly diverted from the centerline of the domain. Such dynamic instability might be due to dynamic contact line that leads to the spatial variation of the contact angles, i.e. contact angle hysteresis. Prescribing static contact angle at the plate would inaccurately capture the rivulet behavior, and therefore dynamic contact model is needed. To explain this behavior, flow simulations were conducted for both dynamic ($\gamma_A = 74.8°$, $\gamma_R = 53.3°$ and $\gamma_E = 70°$) and static ($\gamma_S = 70°$) contact angles at the plate. Given the dynamic nature of this phenomenon, the results presented correspond to a single snapshot in time wherein the experimental results and simulation predictions show similar behavior. As described in the section 2.1, the equilibrium capillary number ($Ca_E$) is one of the input parameters to specify the dynamic contact angle at the solid surface, which needs to be evaluated. Accordingly, the flow simulations were conducted for DCA ($\gamma_A = 74.8°$, $\gamma_R = 53.3°$ and $\gamma_E = 70°$) at a flow rate of $9.30 \times 10^{-7}$ $m^3/sec$ and different $Ca_E$ values. As shown in Figure 6, the predicted shape of the rivulet at $Ca_E = 0.001$ matches well with the experimental shape of the rivulet. Therefore, $Ca_E = 0.001$ was chosen as one of the input parameters for DCA in further flow simulations. Further, the predicted shape of the rivulet has also been compared with the experimental shape for water rivulet at two flow rates ($2.20 \times 10^{-6}$ and $9.30 \times 10^{-7}$ $m^3/sec$) for DCA boundary condition specified at the plate. As shown in Figure 7, the SCA specified at the plate does not precisely predict the motion of the contact line, thereby nature of dynamic behavior of rivulet. The meandering rivulet generally emerges at the low flow rate where surface tension force dominates over other forces. The asymmetrical surface tension forces on either side of the rivulet gives rise to radius of curvature in the meandering rivulet. As seen Figure 7, the radius of curvature of the meandering rivulet found to be increasing with decreased flow rate in both studies. Overall, the predicted shape of the rivulet matches well with the experiment for water at two flow rates.

Indeed, validation of VOF simulation for film and rivulet flow against experiments were reported for various setups [15, 40, 69, 70]. CFD predicted wetted area and film thickness were compare well with the experimental results of Hoffmann et al. [71] and Nusselt theory [72], respectively for film falling over inclined plate [15]. Later, it was also validated against in-house experiment for water, 10 cS and 100 cS standard silicon oil [69]. Further, VOF predicted wetting of a textured corrugated sheet of Montz B1-300 packing qualitatively good agreement with experiments of Subramanian and Wozny [28]. Indeed, VOF method has been successfully used in flow investigation by other for hydrodynamics [10, 15, 40, 69, 70] and reactive mass transfer [57, 73, 74] in two-phase flow. Hence, the VOF method is chosen to study the countercurrent gas–liquid flows in a packed column.

### 4.4 Effects of contact angle

As mentioned earlier, the contact angle is a major component that indicates the degree of wetting in the solid–liquid interaction. A given solvent can show different wetting behavior (i.e., different contact angle) depending on the solid surface. Contact angle hysteresis has also been reported, which is generally considered to be a consequence of either roughness or heterogeneity of the surface [75, 76]. In this view, extensive studies were conducted to understand the effects of static and dynamic contact angles on the interfacial area.

### 4.4.1 Static contact angle

The flow simulations in the REU were restricted to $\gamma_E = 10°$. Effects of contact angle were extensively studied earlier for a rivulet falling over an inclined flat plate [15], corrugated sheet [40], and REU of the Gempak-2A packing [10]. The flow simulations were conducted for 40MEA at a liquid load of $62 \; m^3/m^2h$ and a wide range of contact angles ($\gamma_S \sim 10 - 70°$). Unlike the section 4.2, the pseudo-steady state was decided hereafter when normalized interfacial becomes approximately constant. The net value of the normalized interfacial area ($A_{In}$) at the pseudo-steady was used in further analysis of results. The effect of SCA on $A_{In}$ was briefly studied to show consistency with prior studies.

The effects of contact angle on the normalized effective mass transfer ($A_{eff}$) and interfacial ($A_{In}$) areas for 40MEA (Ka=443) at $q_L = 62 \; m^3/m^2h$ are presented in Figure 8. Both areas are found to initially decrease with increasing γ values. The interfacial area could be the lower value than the effective area. The discrepancy might arise due to the value of mass transfer coefficient used in the computation of effective areas. A structured packed column shows higher value of the overall mass transfer coefficient ($k'_g$) than the corresponding one for wetted wall columns [77, 78]. The theoretical value of ($k'_g$) was considered in the calculation of the effective area, which is much lower than the experimental value. Consequently, a higher value of the effective area is observed. In contrast, the interfacial area consistently decreases with the increased value of the contact angle. The wetting characteristics of a solvent decreases with increased γ value. Consistent with previous studies [15, 40], the interfacial area significantly varies with the contact angle. Further, the value of the interfacial area at $\gamma = 10°$ was found to be three times higher than the corresponding one for $\gamma = 70°$. With decreasing contact angle ($\gamma \leq 40°$), the transition of flow regime gradually occurs, i.e., rivulet to film flow. It is consistent with the previous studies of Haroun et al. [11] and Basden et al [21]. Sebastia-Saez et al. [12] did not observe film flow regime in the simulations, which might be a consequence of boundary conditions in their numerical setup. In case of rivulet flow, the solvent prefers to flow in the channel's valley. At higher $\gamma_S$ values, liquid does not significantly spread and is constrained into the valley [10, 40]. As a result, a thicker rivulet appears that requires very fine mesh to accurately capture mass transfer. A coarser mesh predicts huge value of mass transfer. Subsequently, the effective mass transfer area increases for a fixed theoretical value of $k'_g$. Therefore, the predicted effective area in such condition cannot be accepted. The computation of the mass transfer in such system is limited due to resolving the mesh size in the order of reaction scale. It would be noted that the reaction kinetics would not change the flow morphology. Nevertheless, the CFD predicted interfacial area doesn't have such limitation and shows consistent result with prior reported work. Furthermore, the effect of the contact angle on the interfacial area is found to be pronounced than the previous studies [12, 21, 25]. None the less, Shi and Mersmann [35] and Rocha et al. [42] exaggerated the impact of contact angle on the interfacial area

### 4.4.2 Dynamic contact angle

In the previous section, the static contact angle was considered for sanity check of the consistency of present work with previously reported researches. As mentioned earlier, the instantaneous contact angle differs from the static value. Therefore, DCA was included in the flow simulation to study its impact on the interfacial area. Because of the inadequate grid resolution for reactive flow simulations particularly in partial wetting condition, flow simulations were adhered to hydrodynamics only.

Effects of the DCA on the interfacial area were investigated for NaOH and 40MEA at $q_L = 62 \text{ m}^3/\text{m}^2\text{h}$. The value of the static contact angle was selected as the equilibrium contact angle ($\gamma_E$), i.e., 70° for NaOH. As shown in Figure 1(a), packing unit of Mellapak 250.Y consists of perforated and embossed metal sheets. Sebastia-Saez et al. [79] recently found a lower value of contact angle for the combination of aqueous MEA solvents and textured corrugated sheet of Mellapak 250.X packing. Therefore, the equilibrium contact angle for 40MEA solvent was chosen as 40°. The advancing and receding contact angles were varied around a fixed $\gamma_E$ value for both cases. Next, interfacial areas are compared for SCA and DCA.

Figure 9 (a) shows the temporal evolution of the normalized interfacial area for NaOH at $q_L = 62 \text{ m}^3/\text{m}^2\text{h}$ with $\gamma_A = 80°$ and $\gamma_R = 60°$. As mentioned earlier, physical properties ($\mu$ and $\sigma$) of NaOH are approximately same as those for water (Table 2). The static contact angle for the combination of water and smooth steel surface was found to be 70°, and the contact angle hysteresis can exceed 20° [52]. As expected, periodic oscillation in the normalized interfacial area around its mean value is observed. The oscillation in due to dynamic rivulet as well as breakup of the droplet in the flow domain. On the contrary, the interfacial area was found to achieve the pseudo steady state for static contact angle in the previous studies [10]. To explain this distinct behavior, temporal snapshots of the interfacial area are shown in Figure 9(b). Three flow regimes: droplet, rivulet, and slander liquid threads are seen. Formation of thicker fronts of rivulets is due to capillary forces, which are characterized by higher value of advancing contact angle. Aqueous NaOH solvent is characterized by higher values of surface tension and advancing contact angle; subsequently, the liquid front becomes thicker as time progresses. When liquid rivulet approaches the inflexion plane near trough, rivulet fronts from each sheet meet and coalescence occurs. Eventually, a thick and heavy slander liquid thread appears around 0.25 sec. Further, the bulge of liquid in the open space detached due to gravity and falls on the sheets. Consequently, smaller daughter droplets and rivulet appears (t = 0.41, 0.77 sec). In addition, breakup of small rivulet also gives rise to formation of small droplets units (first unit at t=0.77 and 1.28 sec). The continuous periodic formations of droplets in addition to the presence of stable rivulet in the valley of sheet (t=1.28 and 1.86sec) leads to oscillation in the value of interfacial area. The breakup of liquid rivulet occurs due to end pinching off phenomenon, which has been extensively explained earlier [69]. Subsequently, interfacial area does not achieve a net pseudo steady-state value. The effect of the DCA might not be pronounced as previously found for the rivulet falling over a flat inclined plate. In a flat plate, edge of the rivulet is free and contact line lateral motion is not restricted by the wall. Eventually, A significant lateral movement around the centerline of the inlet for dynamic rivulets can be observed in Figure 7 for both flow rates. On the other hand, the width of the corrugated channel is small, and solvent prefers to flow in the valley of the triangular channel. The contact line movement is constrained by the triangular wall; therefore, the impact of the DCA on the interfacial area is not significantly high.

Next, flow simulations were extended for the combination of different values of advancing and receding contact angles to investigate the impact of contact angle hysteresis on the interfacial area. The value of $\gamma_A$ and $\gamma_R$ are varied around $\gamma_E = 70°$ at $q_L = 62 \text{ m}^3/\text{m}^2\text{h}$. Note that the contact angle hysteresis is associated with the wetting of surface [45, 80]. Figure 10 shows the temporal evolution of $A_{In}$ for these cases. As shown earlier, the normalized interfacial area does not achieve pseudo steady state for DCA. On contrary, using static contact angle leads to a net pseudo steady value of the interfacial area after $t = 0.90$ sec. As expected, the normalized interfacial area decreases with increasing receding contact angle. A decrease in the advancing contact angle corresponds to the increase in the front spread on the solid surface [81]. Change in the interfacial area could be further understood by the effective capillary force for a drop falling over an inclined plate (Figure 2). The longitudinal components of capillary force due to the moving contact line can be computed as $F_C = \sigma w (\cos \gamma_R - \cos \gamma_A)$, where $w$ is width/diameter of the contact area [80, 82]. With decreasing, advancing, or receding contact angle relative to equilibrium contact angle, the value of the capillary force increases in the flow direction.

To explain this behavior, the shapes of the gas–liquid interface at $t = 2 \text{ sec}$ for NaOH are presented in Figure 10 (b) at $q_L = 62 \text{ m}^3/\text{m}^2\text{h}$ and different $\gamma_A$ and $\gamma_R$ values. Three types of the flow patterns: partial film, rivulet, and droplet are observed. Formation of the bulgy front of rivulet is due to capillary forces and is further enhanced due to a higher $\gamma_A$ value. Recall, aqueous NaOH solvent is characterized by the higher surface tension value. With increasing advancing contact angle, the rivulet front becomes thicker, and breakup of the rivulet occurs due to end pinch-off phenomena. Subsequently, small daughter droplets appear. It can be clearly seen in the top unit at higher advancing and receding angles. This process further repeats with time, and oscillation in the temporal evolution of the interfacial area appears in Figure 10(a). In contrary, the interfacial area achieves a net pseudo steady value with a small undulation for the static contact angle.

The temporal evolution of the interfacial for 40MEA is shown in Figure 11. In contrast to NaOH, 40MEA achieves in a net pseudo steady value of the interfacial area. The interfacial area decreases with the increasing value of receding and advancing contact angles. The pseudo steady shape of the interface can be observed in the inset of Figure 11. Recall, the solvent prefers to flow in the valley, and rivulet breakup does not occur. It can be understood by the effect of viscous dissipation. Note that 40MEA has higher viscosity and lower surface tension values as compare to NaOH. For a solvent having lower equilibrium contact angle and higher viscosity, viscous dissipation is significant that damps the contact line motion [80]. In addition, breakup of the rivulet is also inhibited due to dominant of inertia and viscous forces over capillary force. Another studies for the effects of viscosity and surface tension on dynamic instability of rivulet by Schmuki and Laso [83] also found that the meandering is suppressed at high viscosities. Eventually, the liquid rivulet does not undergo dynamic contact line instability, and the interfacial area achieves a net value.

## 4.5　Effect of initial condition

Thus far, flow simulations for the prediction of interfacial area were conducted with an initially dry sheet. Flow simulations with initially wetted plate for liquid falling over an inclined plate were performed in the past [15, 84, 85]. However, these simulations were performed with the static contact angle. The final shape of flow morphology (film, rivulets and droplets) was found to be independent of the initial condition, no matter whether the plate was initially dry or wet. However, the initially wetted plate shows slightly higher value of the interfacial area than the corresponding one started with a dry plate. This can lead to contact line hysteresis for wetting of the surface that was also shown previously [74]. Note that the static contact angle in the simulation can't accurately predict dynamic behavior such as a meandering rivulet (See Figures 6 and 7). Accordingly, flow simulations were conducted with initially wetted sheets, i.e corrugated sheets covered with 0.40 *mm* thick uniform liquid film for both solvents. The input parameters describing DCA for NaOH and 40MEA are ($\gamma_A = 75°$, $\gamma_R = 53°$ and $\gamma_E = 70°$) and ($\gamma_A = 45°$, $\gamma_R = 35°$ and $\gamma_E = 40°$) respectively.

Temporal evolution of the interface for NaOH at $q_L = 62\ m^3/m^2h$, DCA ($\gamma_A = 75°$, $\gamma_R = 53°$ and $\gamma_E = 70°$) and initially wetted sheets is shown in Figure 12(a). For the initially wetted sheets, thickness of the film gradually decreases due to interfacial surface tension and gravity. The film rupture starts near the crest of triangular channels. i.e at the inflexion region ($t=0.02$ sec). Subsequently, the ruptured film shrinks and get narrowed because of the capillary effect force caused by interfacial surface tension. Further shrinkage of the ruptured film lead to formation of the rivulet which can be seen at $t=0.30$ sec. Later behavior of rivulet flow is found to be similar as those found for the simulation started from initially dry sheets (see for t=0.5 and 1 sec in Figure 12(a)). Although they started from different initial wetting conditions ($t = 0$ sec), flow morphology at the pseudo-steady state is appeared to be same in both cases (see snapshot of the interface at $t = 1.0$ sec in Figure 12(a) and (b)). The case initially started with wetted sheets shows slightly higher value of $A_{In}$ in Figure 12(c), however the difference is not significant. Similar observation is also found in the inset of Figure 12(c) for other DCA values ($\gamma_A = 85°$, $\gamma_R = 60°$ and $\gamma_E = 70°$) at the same liquid load. For the falling film over an inclined plate, initially wetted sheets show higher value of wetted area than the corresponding one started from dry conditions [46, 84]. This might be due to occurrence of hysteresis in the DCA and might have some influence on the wetting. Wetting hysteresis is seen for both cases when they achieve pseudo steady state. The numerical simulation was further carried out at low liquid load $q_L = 31\ m^3/m^2h$ to investigate the effect of inertia of the wetting hysteresis. The difference in the interfacial area was computed as $\Delta A_{In} = (A_{In,W} - A_{In,D})/A_{In,W}$, where $A_{In,W}$ and $A_{In,D}$ are the normalized interfacial areas for the initially wetted and dry sheets respectively to present wetting hysteresis. As expected, a slightly higher value of $\Delta A_{In}$ was found at lower liquid load when solvent achieves pseudo steady state (see Figure 12(d)). This is somewhat expected due to the enhanced contact line mobilization at low inertia. Inertia tends to retard the contact line movement [86]. Wetting hysteresis is not observed for 40MEA solvent in Figure 13. Both show almost the same value of the interfacial area at $t \geq 1.0$ sec. Inset of Figure 13 shows an identical shape of the rivulet along with small droplets. The effect of the contact line hysteresis in minimal for a solvent having smaller $\gamma_E$ and $\sigma$ values. Subsequently, it leads to nearly same value of the normalized interfacial area regardless initial conditions in such system.

## 5. Conclusion

The CFD modeling for the hydrodynamics of countercurrent flows in a structured packed column is multiscale problem. 3-D multiphase flow simulations using the VOF method were conducted to explore local flow behavior in the Mellapak 250.Y packings. The impact of liquid loads, dynamic contact angle, and initial wetting condition of the sheets on the interfacial area were extensively explored. Note that the overall mass transfer is dominated by chemical kinetics in which the interfacial area plays a key role.

The CFD predicted effective area via Danckwerts method reasonably matches with corresponding experimental values [19] at a wide range of liquid loads. As expected, the interfacial area initially increases with increased liquid load and then remains unaffected. The effect of the surface characteristics on the wetting and spreading is studied via contact angle boundary condition at the corrugated sheets. Contact angle is one of the critical factors that dictate the wetting, thereby the interfacial area. The CFD predicted shape of the meandering rivulets was compared to the experimental observations of rivulet falling over an inclined plate. Both results show a reasonably good agreement for dynamic contact angle specified at the plate surface. Further, the effects of both dynamic and static contact angles boundary condition at the sheets were explored. The interfacial area increases with decreasing value of the static contact angle at a fixed flow rate for a given solvent. To explore the effects of contact angle hysteresis on the interfacial area, the DCA was specified at the sheet. The contact angle hysteresis has pronounced impact on the interfacial area for a solvent possessing higher surface tension value, thereby higher equilibrium contact angle. The interfacial area shows temporal oscillation due to formation of small discontinuous rivulets and droplets. Furthermore, the normalized interfacial area does not achieve the pseudo steady state. In contrast, the interfacial area gets a net pseudo steady value for a solvent having low surface tension. The effect of the initial sheet conditions (dry vs wet) is also studied at two liquid loads and solvents. An initially wetted sheet shows a slightly higher value of interfacial area as compared to initially dry sheet for 0.1M NaOH at a fixed liquid load. Lower value of wetting hysteresis is observed at a higher liquid load due to impact of inertia on contact line motion. On the other hand, 40MEA does not shows wetting hysteresis which might be due to high viscosity and low surface tension. The outcome would be helpful for fundamental understanding of the local hydrodynamics. The interfacial area model can be improved and integrate into device scale modelling for the efficient packing design.

**Acknowledgments**

Pacific Northwest National Laboratory is operated by Battelle for the U.S. Department of Energy (DOE) under Contract No. DE-AC05-76RL01830. This work was funded by the DOE Office of Fossil Energy's Carbon Capture Simulation Initiative (CCSI) through the National Energy Technology Laboratory. Authors also acknowledge the valuable inputs and discussions of Prof. Gerry Rochelle (University of Texas at Austin).

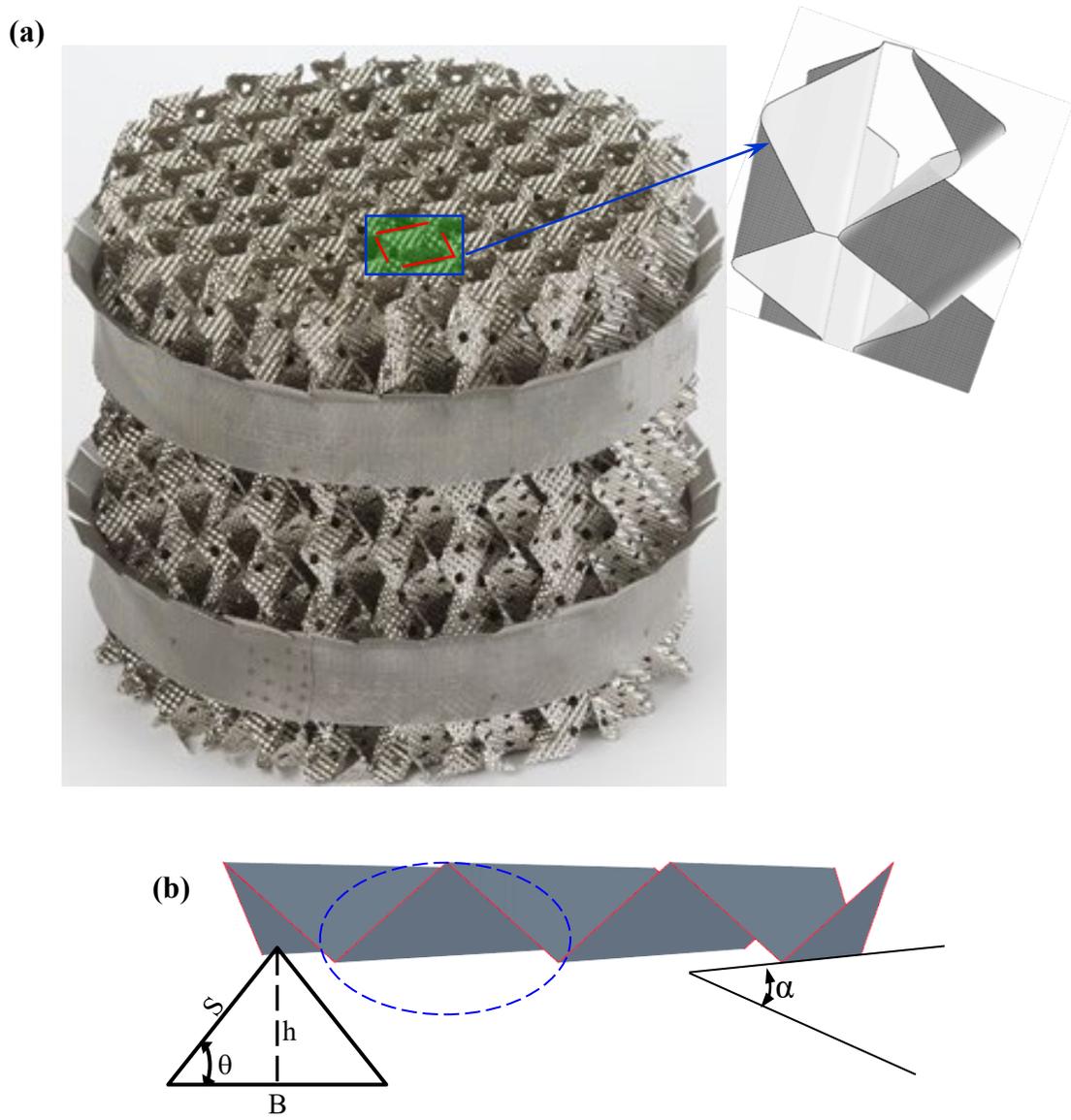

**Figure 1:** (a) An image of a structured packing shows the perpendicular arrangement of corrugated sheets. Exploded view describes the selection of current REU model in the flow simulations (Source: https://www.sulzer.com). (b) Design of the corrugated sheet to construct the model of REU of Mellapak 250.Y packing.

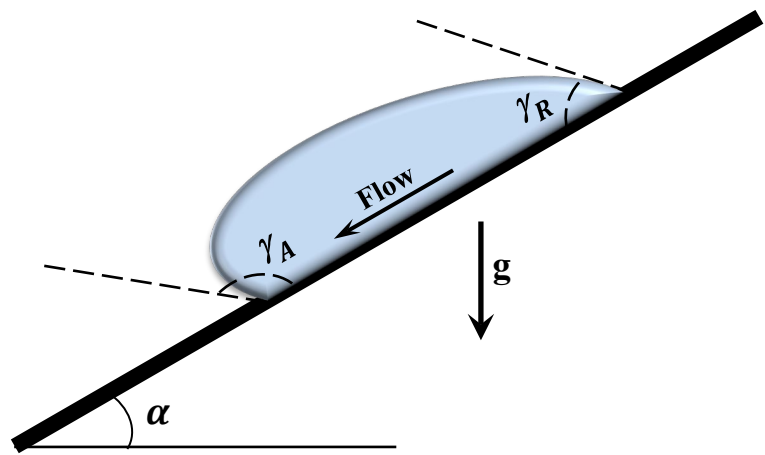

**Figure 2:** A liquid droplet falling over an inclined flat plate demonstrates dynamic [advancing ($\gamma_A$) and receding, ($\gamma_R$)] contact angles.

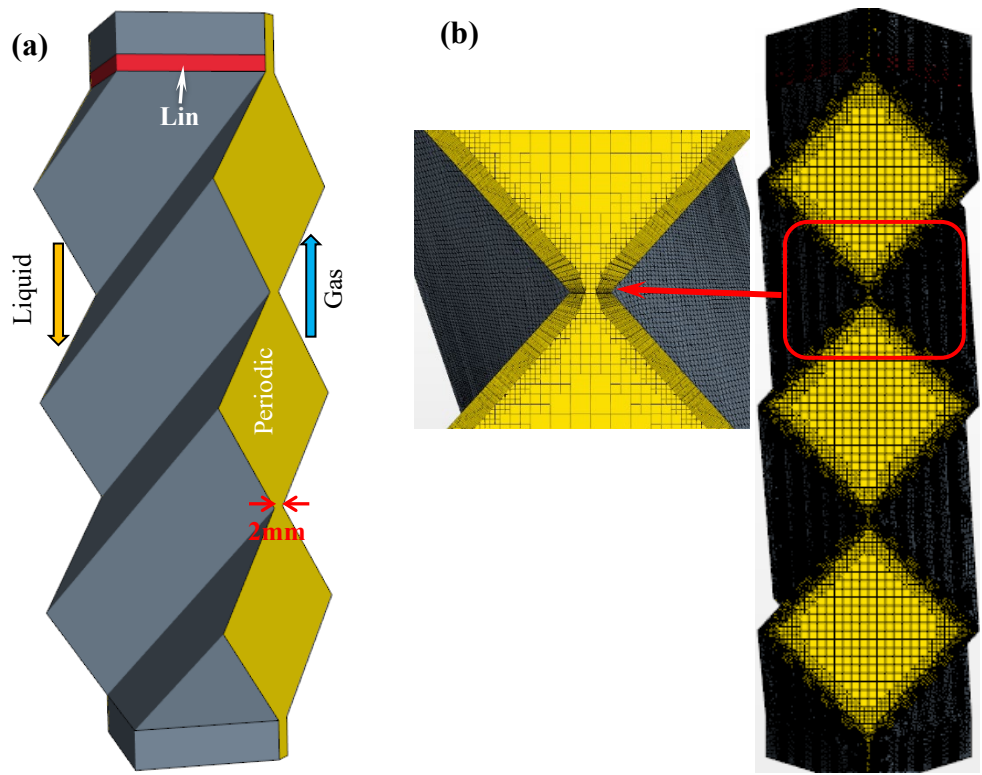

**Figure 3:** (a) Three-dimensional view of the flow domain shows boundary conditions employed and the direction of gas and solvent in the countercurrent flows. The periodic boundary condition is specified at the lateral sides of the domain. The solvent was fed along the inner perimeter of the top and the gas was injected from the bottom of the flow domain. (b) Discretized flow domain shows the rectangular nonuniform mesh in flow domain and very fine mesh near the sheets where gas-liquid interface is expected (exploded view in the center).

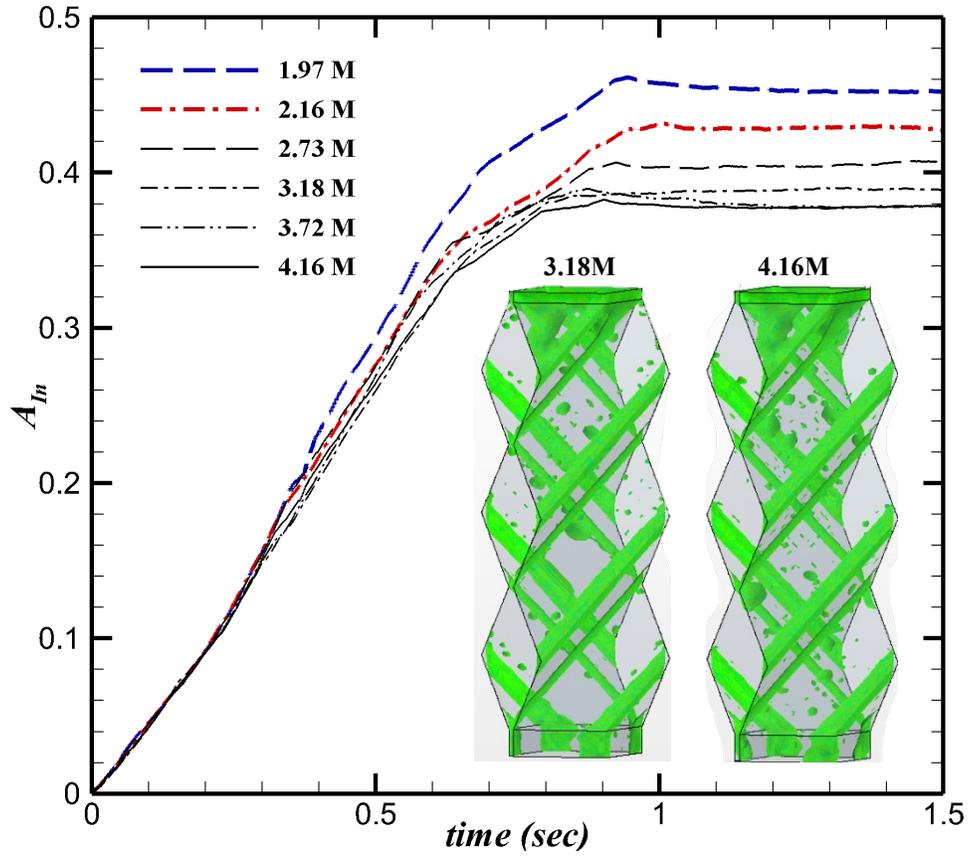

**Figure 4:** Temporal evolution of the normalized interfacial area ($A_{In}$) for 40MEA with DCA ($\gamma_A = 45°$, $\gamma_R = 30°$ and $\gamma_E = 40°$) at $q_L = 62\ m^3/m^2 h$ and six grid resolutions. Inset shows identical pseudo-steady shape of the interface at the grid resolutions 3.18M and 4.16M.

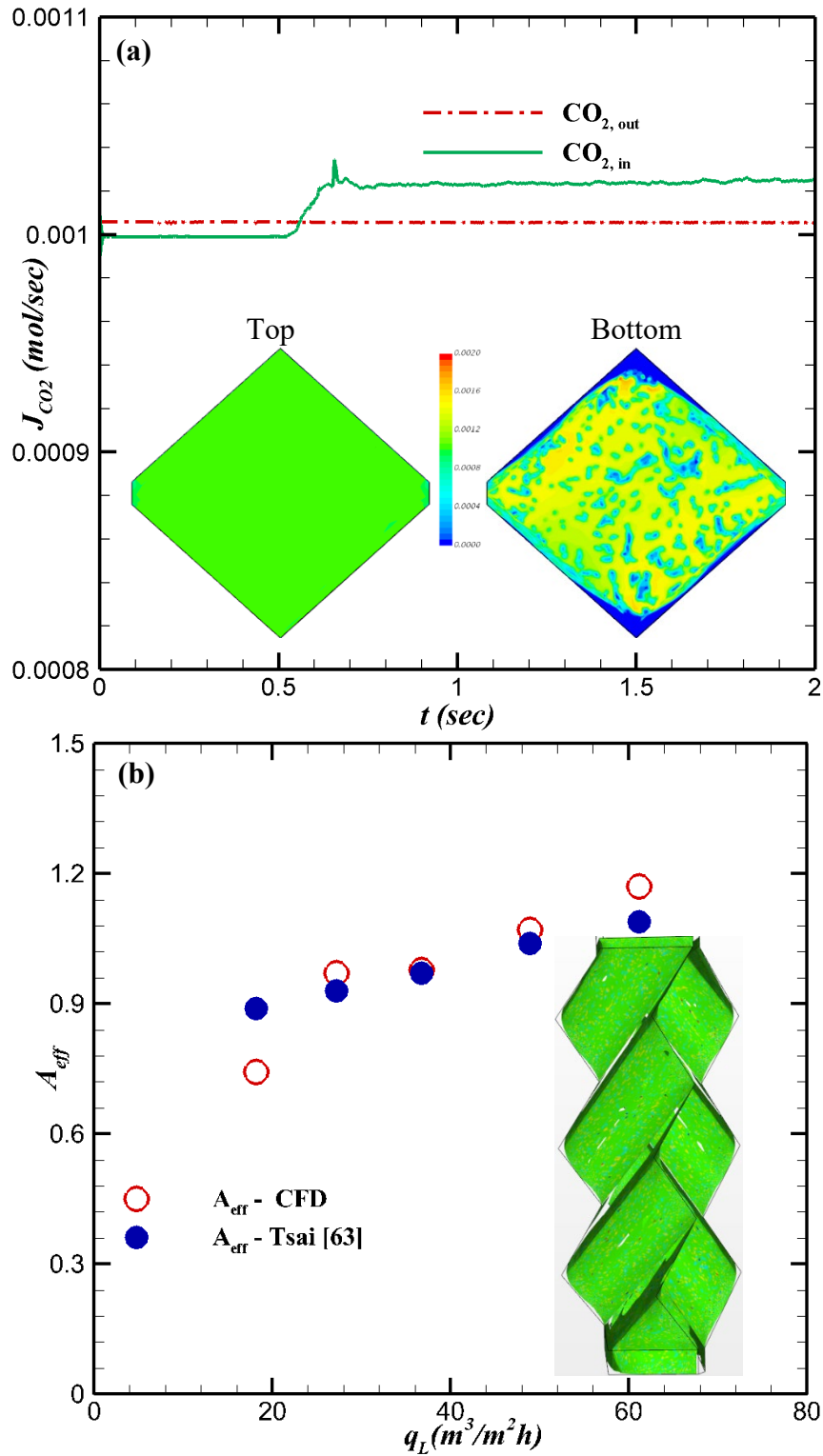

**Figure 5:** (a) Temporal variation of the $CO_2$ flux at the gas inlet and the outlet. The pseudo-steady state was assumed when $CO_2$ flux at these boundaries achieves nearly constant values. Inset shows the contours of $CO_2$ flux at the gas inlet and outlet. (b) Comparison of the CFD predicted effective area with corresponding experimental value (Tsai [63]) at different liquid loads. Inset shows the snapshot of the gas-liquid interface at $q_L = 62\ m^3/m^2h$.

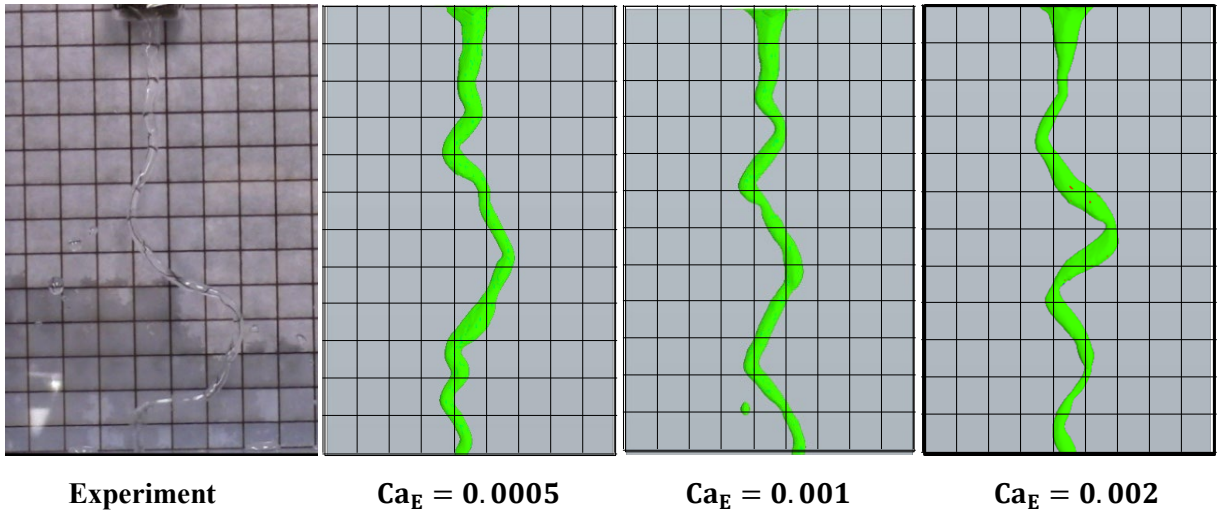

**Figure 6:** Comparison between experimentally observed and CFD predicted shapes with DCA ($\gamma_A = 74.8°$, $\gamma_R = 53.3°$ and $\gamma_E = 70°$) for a water rivulet falling over an inclined plate at a flow rate $9.3 \times 10^{-7}$ m$^3$/sec and different $Ca_E$ values. The meandering rivulet appears due to the dynamic contact line instability.

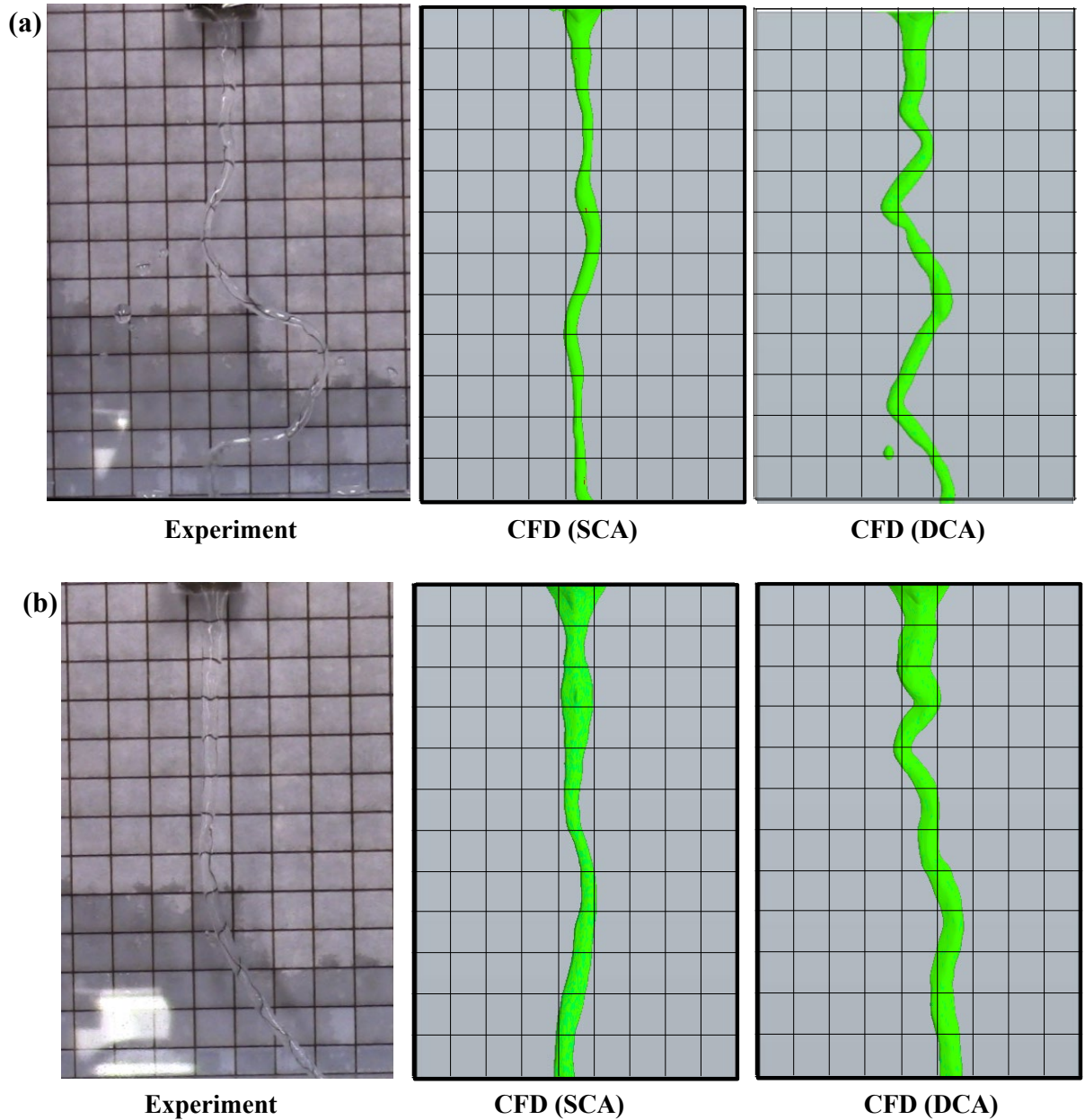

**Figure 7:** Comparison of the CFD predicted shapes using DCA ($\gamma_A = 74.8°$, $\gamma_R = 53.3°$ and $\gamma_E = 70°$), SCA ($\gamma_S = 70°$) and experimental shapes for water rivulet falling over an inclined plate at two flow rates (a) $2.2 \times 10^{-6}$ and (b) $9.3 \times 10^{-7}$ m³/sec. The dynamic instability in rivulet appears at both flow rates.

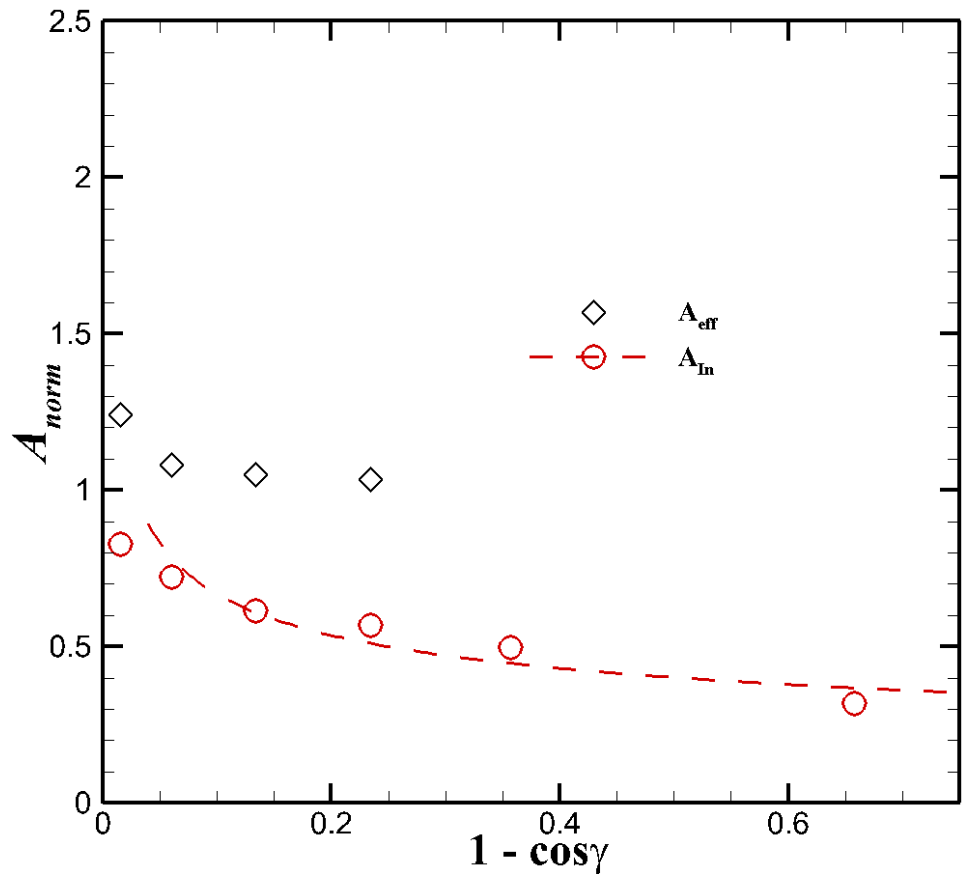

**Figure 8:** Variation of the normalized effective ($A_{eff}$) and interfacial ($A_{In}$) areas with SCA ($\gamma_S$) for 40MEA at $q_L = 62 \; m^3/m^2 h$. Both decrease with increasing value of SCA.

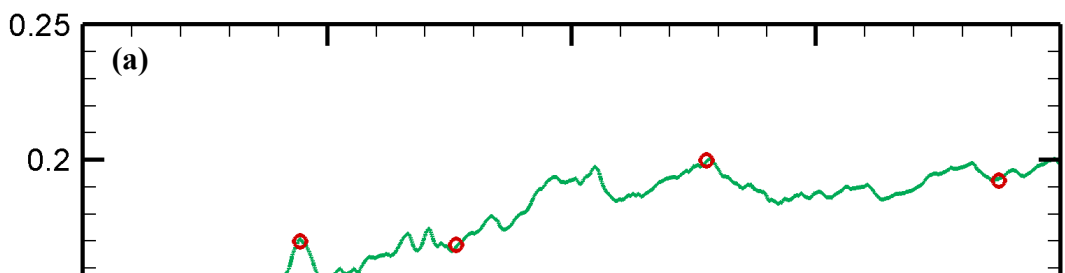

(a)

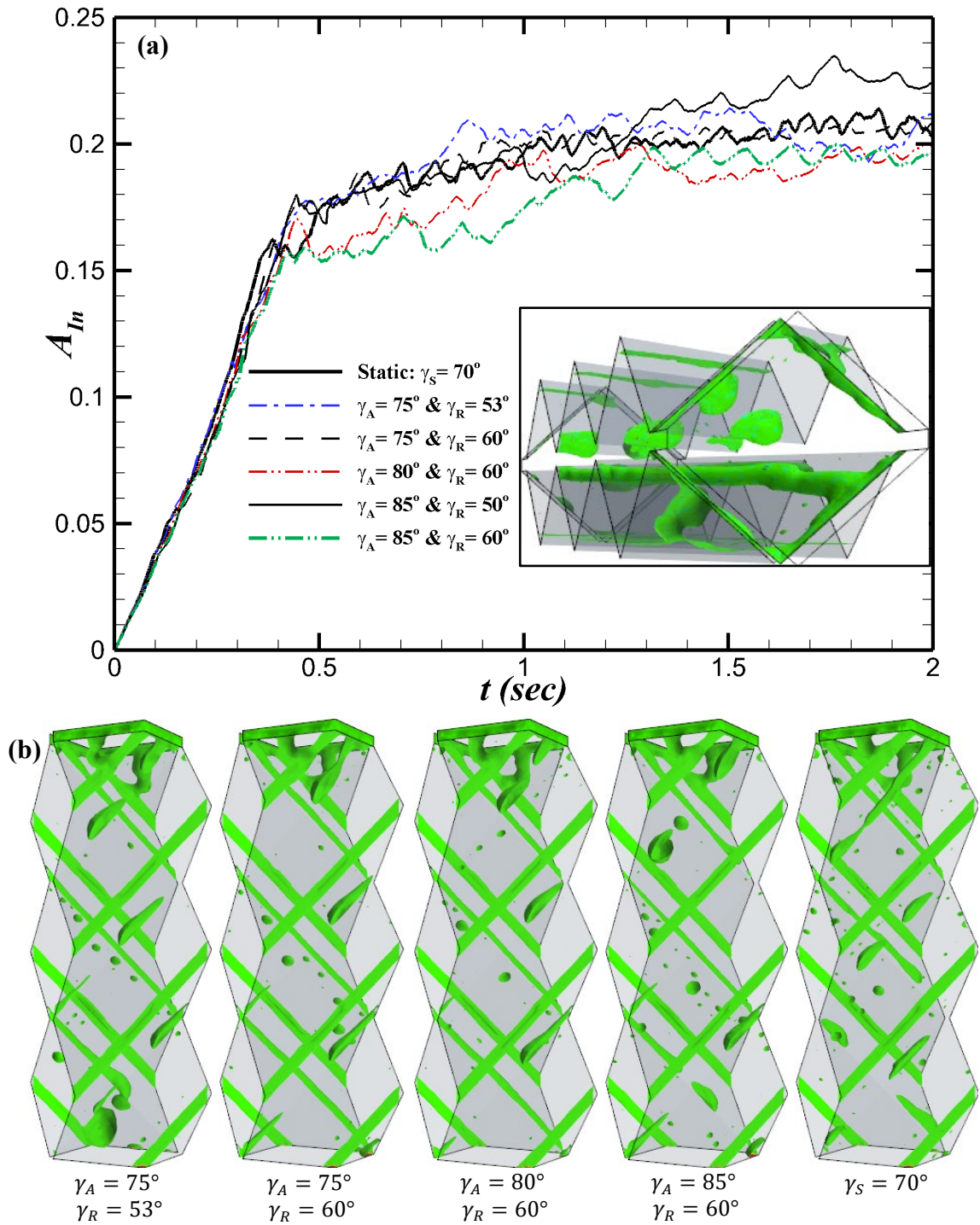

**Figure 10:** (a) Temporal evolution of $A_{In}$ for NaOH at $q_L = 62 \ m^3/m^2 h$, $\gamma_E = 40°$ and different $\gamma_A$ and $\gamma_R$ values. Inset shows that solvent is stuck to the respective sheets. (b) Shape of the interface in the REU for the same cases at $t$ = 2.0 sec.

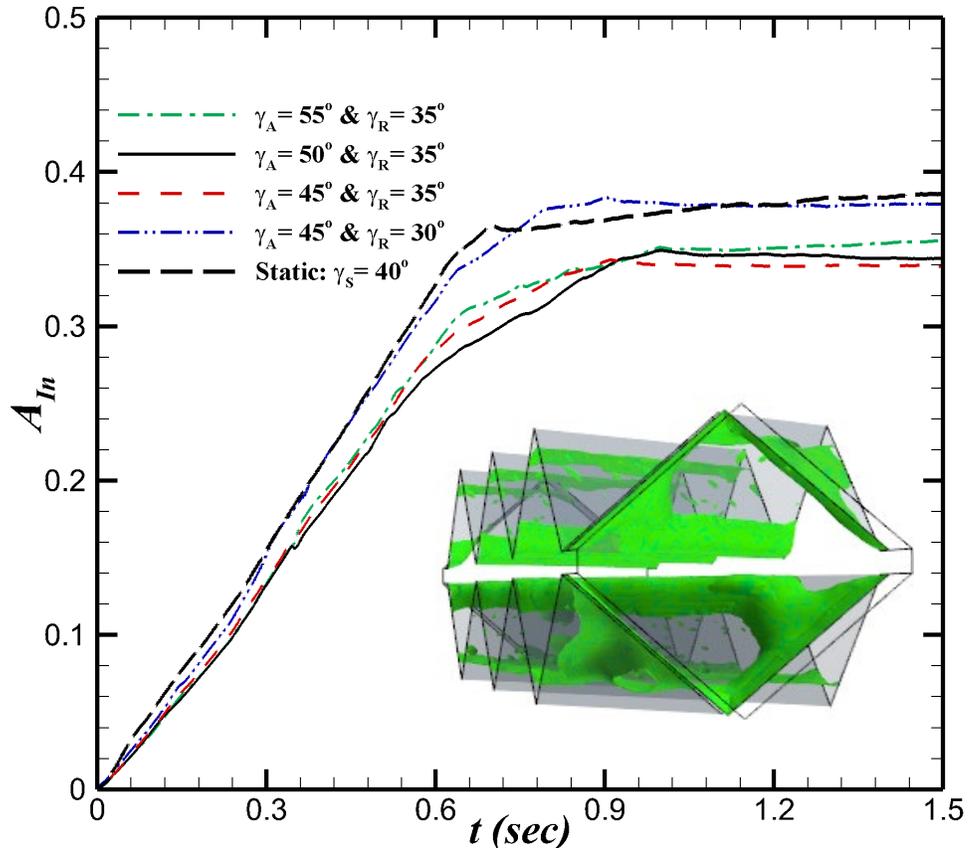

**Figure 11:** Transient evaluation of $A_{In}$ for 40MEA at $q_L = 62\ m^3/m^2h$, $\gamma_E = 40°$ and different $\gamma_A$ and $\gamma_R$ values. Inset shows the top view of gas-liquid interface in the REU for the same flow rate at $\gamma_A = 45°$ and $\gamma_R = 30°$.

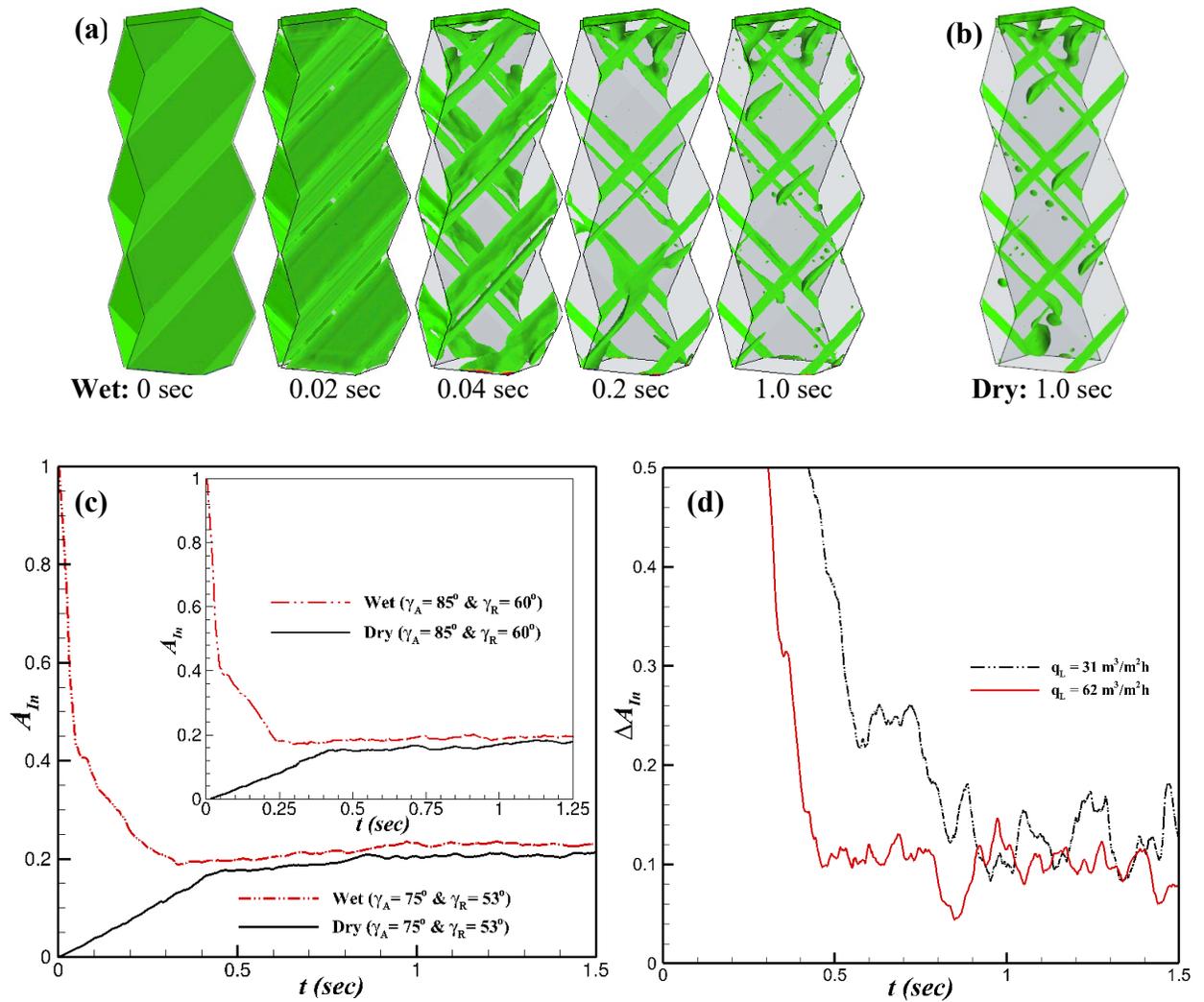

**Figure 12:** (a) Evolution of the shapes of interface for NaOH at $q_L = 62\ m^3/m^2h$, DCA ($\gamma_A = 75°$, $\gamma_R = 53°$ and $\gamma_E = 70°$) and initially wetted sheets (b) Snapshot for the interface at same $q_L$ and DCA values except initially dry sheets. (c) Comparison of the temporal value of $A_{In}$ for initially wetted and dry sheet. Inset shows the same plot at DCA ($\gamma_A = 85°$, $\gamma_R = 60°$ and $\gamma_E = 70°$). (d) Plot shows change of wetting is slightly higher at the low liquid load (i.e. lower inertia) at the same DCA value.

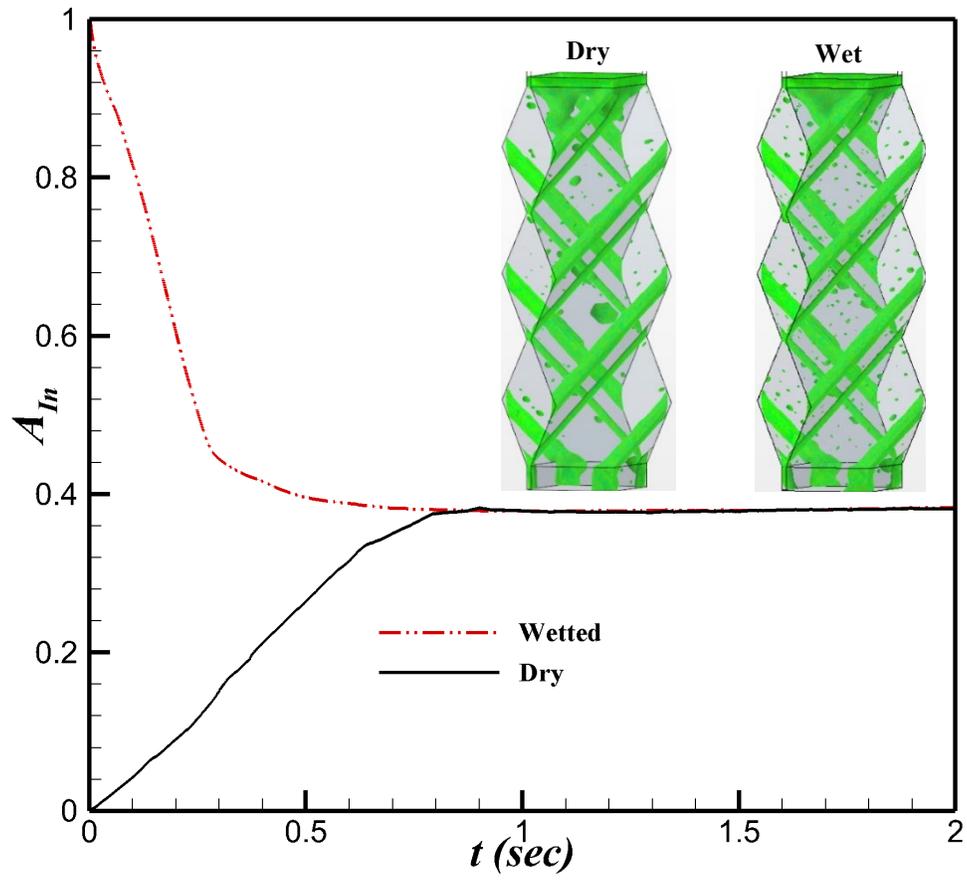

**Figure 13:** Comparison of $A_{In}$ for initially wetted and dry sheets for 40MEA at $q_L = 62\ m^3/m^2 h$ and DCA ($\gamma_A = 45°$, $\gamma_R = 30°$ and $\gamma_E = 40°$. Inset shows the pseudo-steady shape of the interface at both cases.